\documentclass{mn2e}

\usepackage{amsfonts}
\usepackage{amsmath}
\usepackage{amssymb}
\usepackage{graphicx}

\newcommand{\Cal}[1]{\ensuremath{\mathcal{#1}}}
\newcommand{\I}[1]{\ensuremath{\Cal{I}_{#1}}}
\newcommand{\F}{\Cal{F}}
\newcommand{\del}{\ensuremath{\delta}}
\newcommand{\delc}{\ensuremath{\delta_{c}}}
\newcommand{\delv}{\ensuremath{\delta_{v}}}
\newcommand{\delT}{\ensuremath{\delta_{T}}}
\newcommand{\nuc}{\ensuremath{\nu_{c}}}
\newcommand{\nuv}{\ensuremath{\nu_{v}}}
\newcommand{\nuT}{\ensuremath{\nu_{T}}}
\newcommand{\Fsw}{\ensuremath{\Cal{F}_{\rm SvdW}}}

\newcommand{\erfc}[1]{\ensuremath{{\rm erfc}\left(#1\right)}}

\newcommand{\etal}{\emph{et al}.}

\newcommand{\eqn}[1]{equation~\eqref{#1}}

\newcommand{\fig}[1]{Figure~\ref{#1}}

\newcommand{\ph}[1]{\phantom{#1}}

\newcommand{\be}{\begin{equation}}
\newcommand{\ee}{\end{equation}}

\title[Void hierarchy]
      {A hierarchy of voids:  More ado about nothing}

\author[A. Paranjape et al.]
{Aseem Paranjape$^1$\thanks{E-mail: aparanja@ictp.it},  
 Tsz Yan Lam$^{2}$\thanks{E-mail: tszyan.lam@ipmu.jp},  
 \& Ravi K. Sheth$^{1,3}$\\
 $^1$ The Abdus Salam ICTP, Strada Costiera 11, 34151 Trieste, Italy\\
 $^2$ IPMU, University of Tokyo, Kashiwa, Chiba 277-8583, Japan\\
 $^3$ Center for Particle Cosmology, University of Pennsylvania, 
      209 S. 33rd St., Philadelphia, PA 19104, USA}
\begin{document}
\pagerange{\pageref{firstpage}--\pageref{lastpage}}

\maketitle 

\label{firstpage}

\begin{abstract}
We extend earlier work on the problem of estimating the void-volume 
function -- the abundance and evolution of large voids which grow 
gravitationally in an expanding universe -- in two ways.  The first 
removes an ambiguity about how the void-in-cloud process, which erases 
small voids, should be incorporated into the excursion set approach.  
The main technical change here is to think of voids within a fully 
Eulerian, rather than purely Lagrangian, framework.  The second 
accounts for correlations between different spatial scales in the 
initial conditions.  We provide numerical and analytical arguments 
showing how and why both changes modify the predicted abundances 
substantially.  In particular, we show that the predicted importance 
of the void-in-cloud process depends strongly on whether or not one 
accounts for correlations between scales.  With our new formulation, 
the void-in-cloud process dramatically reduces the predicted abundances 
of voids if such correlations are ignored, but only matters for the 
smallest voids in the more realistic case in which the spatial 
correlations are included.  
\end{abstract}

\begin{keywords}
large-scale structure of Universe
\end{keywords}

\section{Introduction}
The abundance of clusters and its evolution is a useful probe of 
the primordial fluctuation field, the subsequent expansion history 
of the universe, and the nature of gravity.  
This is, in part, because there is an analytic framework for 
understanding how cluster formation and evolution depends on the 
background cosmological model (Gunn \& Gott 1972; Press \& Schechter 1974;
Peacock \& Heavens 1990; Bond \etal\ 1991; Sheth, Mo \& Tormen 2001; 
Martino, Stabenau \& Sheth 2009).  

If the clusters were identified in a galaxy survey, then it is 
possible to identify underdense regions -- voids -- in the same 
dataset (e.g. Kauffmann \& Fairall 1991; Hoyle \& Vogeley 2004; 
Hoyle \etal\ 2005; Patiri \etal\ 2006; Pan \etal\ 2011).  
As for the clusters, there exists an analytic framework for 
understanding void formation (
Blumenthal \etal\ 1992; Dubinski \etal\ 1993; 
van de Weygaert \& van Kampen 1993; Sheth \& van de Weygaert 2004; 
Furlanetto \& Piran 2006; Patiri, Betancort-Rijo \& Prada 2006), 
so the comoving number density of voids of radius $R$, and its 
evolution, provides complementary information about cosmology 
(Kamionkowski, Verde \& Jimenez 2009; Lam, Sheth \& Desjacques 2009; 
D'Amico \etal\ 2011) and gravity (Martino \& Sheth 2009).  
Void shapes are interesting too 
(Park \& Lee 2007; Biswas, Alizadeh \& Wandelt 2010; Lavaux \& Wandelt 2010), 
but they are not the primary interest of this paper.  

Following Press \& Schechter (1974), studies of cluster and void 
evolution relate the formation of an object to its initial overdensity.  
A cluster today is a region that is about 200 times the background 
density, and it formed from the collapse of a sufficiently overdense 
region in the initial conditions.  
However, the overdensity associated with a given position in space 
depends on scale (in homogeneous cosmologies, the likely range of 
overdensities is smaller on large scales).  So, to estimate cluster 
abundances, the problem is to find those regions in the initial 
conditions which are sufficiently overdense on a given smoothing scale, 
but not on a larger scale.  This is because, if the larger region is 
sufficiently overdense, then, as it pulls itself together against the 
expansion of the background universe and collapses, it will also squeeze 
the regions within it to smaller and smaller sizes.  The framework 
for not double-counting the smaller overdense clouds that are embedded 
in larger overdense clouds is known as the Excursion Set approach 
(Epstein 1983; Bond \etal\ 1991; Lacey \& Cole 1993; Sheth 1998).

For voids -- regions that today are about 20\% the background density -- 
the problem is slightly more complicated, since one must account not 
just for the analogous void-in-void problem, but also for the fact 
that underdensities which are surrounded by sufficiently overdense 
shells will be crushed as the overdensity collapses around them.  
This void-in-cloud problem was identified by 
Sheth \& van de Weygaert (2004), who also showed how one might 
account for both the void-in-void and the void-in-cloud problems in 
the language of the excursion set approach.  

However, their formulation suffers from an important drawback -- they 
treat the identification of the overdensity associated with a cloud as 
a single scale independent number.  As they noted, it is easy to see 
that this is, at best, a crude approximation.  Suppose that this number 
is that associated with the formation of a cluster.  Then, their approach 
corresponds to eliminating from the list of all possible voids all those 
that are surrounded by an initially larger region which is destined to 
have collapsed and formed a cluster by the time the void they surround 
would have formed (were it not surrounded by this overdensity).  
This leads to the question of what to do with sufficiently underdense 
regions which were surrounded by regions which will not have collapsed 
completely by the time the void inside them forms, but that will 
nevertheless have squeezed the enclosed void, thus altering its size, 
and possibly even preventing its formation.  

To illustrate the magnitude of this effect, Sheth \& van de Weygaert 
showed how the predicted void abundances change if one uses the 
overdensity associated with `cloud' turnaround instead of collapse 
(the two differ by approximately a factor of 1.6 in initial overdensity).  
While the difference for big voids is small -- big underdense regions 
are unlikely to be surrounded by even larger overdensities -- the effect 
on smaller voids is dramatic.  Although it is the largest voids which 
are most easily measured, and so most likely to place the most interesting 
constraints on cosmological models, the uncertainty from not knowing 
precisely where the void-in-cloud problem becomes relevant is problematic.  
(One way to view this problem is to note that large voids are exponentially 
rare, so to constrain cosmology requires large survey volumes.  By 
having an accurate model of smaller voids, one potentially allows 
smaller surveys to place interesting constraints.)  

The main goal of the current paper is to present a formulation of 
the problem which resolves this drawback of the initial formulation.  
The key is to phrase the criterion for being a void in terms of 
the late time field -- a void is the largest region in the late 
time field which is sufficiently underdense -- and to then determine 
what this requires of the initial field.  This means one must be 
able to relate what are often called Eulerian volumes in the 
late-time field, and Lagrangian ones in the initial field.  
Fortunately, this can be done within the Excursion Set approach 
(Sheth 1998; Lam \& Sheth 2008).  

However, there turn out to be a number of subtleties along the way, 
which are related to one of the technical assumptions associated 
with the Excursion set approach.  Strictly speaking, the overdensity 
associated with a given position and scale is correlated with the 
overdensity on all other smoothing scales as well.  Therefore, if one 
plots this overdensity as a function of smoothing scale, then this 
looks like a random walk with correlated steps.  
Following Bond \etal\ (1991), most excursion set analyses, and the 
Sheth \& van de Weygaert model for voids in particular, make the 
approximation that the steps are uncorrelated, and they then assume 
that the resulting prediction will be a useful approximation to that 
which one would have obtained if one had solved the (more physically 
relevant) correlated steps problem.  Accounting for such correlations 
makes relatively minor changes to the cloud-in-cloud (or void-in-void) 
predictions (Peacock \& Heavens 1990; Maggiore \& Riotto 2010; 
Paranjape et al. 2011).  In what follows, we will show that the 
difference between the correlated and uncorrelated solutions is 
much larger for the void-in-cloud problem.  

In Section~\ref{constant+linear} we show how to cleanly resolve 
the void-in-cloud issue in the Excursion set approach.  
In Section~\ref{monteCarlos}, we use a numerical Monte-Carlo method 
to show that the uncorrelated steps formulation is quite sensitive 
to this change -- our solution to the void-in-cloud problem 
predicts far fewer large voids than do Sheth \& van de Weygaert.  
On the other hand, the correlated steps formulation is almost 
completely unaffected by the void-in-cloud problem in the first 
place, and so is not sensitive to the change in our prescription.  
In other words, the void-in-cloud problem is a case in which the 
difference between correlated and uncorrelated steps matters greatly. 
A final section discusses some implications.

\section{A better model of the void-in-cloud problem}\label{constant+linear}
In this section, we show how a more careful statement of the void-in-cloud 
process leads to a slightly modified formulation of the problem in the 
Excursion set approach.  In essence, this resolution of the problem 
combines the analysis in Sheth \& van de Weygaert (2004) with that in 
Sheth (1998).  

\subsection{Lagrangian vs Eulerian treatments}
In what follows, we will denote the Eulerian radius and volume of 
the void by $R$ and $V$ respectively (so that $V=4\pi R^3/3$), and 
refer to Lagrangian length scales simply through the associated mass 
$m = \bar\rho (4\pi R_{\rm L}^3/3)$, where $\bar\rho$ is the comoving 
background density and $R_{\rm L}$ is the Lagrangian radius which
evolved into the Eulerian radius $R$.  We will also use $s(m)$ to denote
the variance of the linearly extrapolated density contrast when
filtered on a Lagrangian scale corresponding to mass $m$: $s(m) =
(2\pi^2)^{-1} \int_0^{\infty}{\rm d}k \,k^2 P(k) W^2(k R_{{\rm L}})$,
where $W(kR_{\rm L})$ is the filter and $P(k)$ the linearly evolved
matter power spectrum.

The condition for being identified as a void of Eulerian size $R$ 
at some time $t$ is that the region of size $R$ must be 
 (a) less dense than some critical threshold (typically about twenty 
     percent of the background density);
 (b) denser than this critical threshold value on all larger 
     Eulerian scales.
Sheth \& van de Weygaert replaced these Eulerian conditions with 
Lagrangian ones.  The Lagrangian region of mass scale $M$ must be 
 (a$_{\rm L}$) less dense than some critical density initially (typically, 
      linear theory overdensity of $-2.71$); 
 (b$_{\rm L}$) denser than this on all larger mass scales; and 
 (c$_{\rm L}$) not dense enough on these larger Lagrangian scales for this 
     to have influenced the evolution of the initial void-candidate 
     region sufficiently that it did not form a void at late times.

Sheth \& van de Weygaert argued that these requirements correspond to 
two different barriers in the Excursion set approach.  In the plane 
of (linearly extrapolated) initial overdensity versus scale, the first 
two requirements correspond to the first crossing of a barrier of 
constant height \delv.  (When extrapolated to the present time using 
linear theory, $\delv=-2.71$, approximately independent of the 
background cosmology, and this fixes the void mass as 
$M\approx0.2\bar\rho V$, see below.)  At issue is how best to 
implement the last constraint (c$_{\rm L}$):  i.e., how to remove from 
the list of potential voids identified in the initial conditions, 
those which would not also be identified as voids at later times 
(i.e. in the Eulerian field).  

Sheth \& van de Weygaert assumed that this could be done simply by 
introducing a second barrier, $B$:  
Of the set of walks which first cross \delv\ at the Lagrangian
scale corresponding to the void mass $M$, one must remove those which 
crossed $B$ before (i.e. at some mass $m>M$) they crossed \delv. 
They assumed that $B=\delc=\,$constant (and hence parallel to \delv),
where \delc\ is the initial overdensity required for collapse at some 
time $t$, extrapolated using linear theory to time $t$ 
(if $\delta_v=-2.71$ then $\delta_c=1.686$)\footnote{Strictly speaking, 
the exact values of \delc\ and \delv\ depend weakly on the cosmological 
parameters; the values quoted above correspond to the 
Einstein-deSitter case. Our focus, however, is on conceptual issues 
rather than numerical accuracy, and our discussion does not depend on 
the exact numerical values of these parameters.}. 
This would correspond to excluding regions which surround the void 
candidate region and have collapsed by time $t$, thus completely 
squeezing out the void.

In our new approach, which allows us to account for regions which 
have only partially squeezed the void (and were not excluded by 
Sheth \& van de Weygaert), it turns out to be more straightforward 
to not mix-and-match conditions in Eulerian and Lagrangian space.  
Rather, we will work entirely with the conditions (a) and (b) stated 
in Eulerian space, and we will draw on the analysis in Sheth (1998) 
to implement these Eulerian conditions in the (essentially Lagrangian) 
plane of initial overdensity versus scale.  

\begin{figure}
 \centering
 \includegraphics[width=\hsize,height=0.25\vsize]{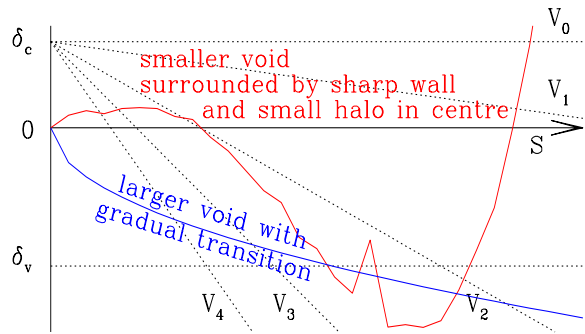}
 \caption{Excursion set model of voids.  
          Dotted lines show the `barriers' associated with Eulerian 
          volumes $V_4>V_3>V_2>V_1>V_0$:  barriers for larger volumes 
          fall more steeply.  
          The horizontal line at $\delta_v=-2.71$ shows the critical 
          linearly extrapolated initial density for voids.  Solid
          lines show two examples of random walks; both first-cross
          $\delta_v$ on the same mass scale $S$.  
          Since neither walk crossed $\delta_c$ prior to 
          crossing $\delta_v$, Sheth \& van de Weygaert would have 
          assigned the same void mass and Eulerian volume to both walks.  
          However, in our prescription, the (blue) one which falls
          monotonically with $S$ is associated with a larger Eulerian
          volume  (between $V_3$ and $V_2$), because its evolution is
          not modified by the void-in-cloud process:  we would have  
          assigned the same mass and volume to it as they did.  
          The other (red) walk represents an overdensity on the
          Eulerian scale $V_2$ (because $V_2$ is first crossed at
          $\delta>0$), but a void on the Eulerian scale just smaller
          than this (because the first crossing of the next shallower
          barrier will be at $\delta<\delta_v$).  
          The evolution of this void has been modified by the collapse 
          of the overdensity surrounding it.  We would assign a larger 
          mass to the `wall' which surrounds the void, and a smaller 
          mass and volume to the void itself, compared to Sheth \& van 
          de Weygaert.  Moreover, note that, for this walk, the first 
          crossing of \delv\ is actually not so significant.}
 \label{eulVoids}
\end{figure}

\subsection{The Eulerian treatment}
To begin, one notes that the spherical evolution model relates the 
Eulerian overdensity 
 $\Delta_{\rm NL} \equiv m/(\bar\rho V)$, 
where $m$ is the mass in a region that has volume $V$ at time $t$, 
to the linearly extrapolated density contrast $\delta(t)$ by 
\begin{equation}
\Delta_{\rm NL}(t) = \frac{m}{\bar\rho V} 
           \approx \left(1 - \frac{\delta(t)}{\delta_c} \right)^{-\delta_c}
\label{DeltaNL}
\end{equation}
(Bernardeau 1994).  
If $V = 4\pi R^3/3$ is specified, then this relation defines a curve 
$B_V(m)$ which 
gives the value of the linearly extrapolated density contrast in a
Lagrangian region containing mass $m$ which evolves into the Eulerian
volume $V$ at time $t$:
\begin{equation}
 B_V(m) = \delc\left[1 - \left(\frac{m}{\bar\rho V}\right)^{-1/\delc}\right]\,.
\label{BVm}
\end{equation}
Notice that $B_V(m)\to \delta_c$ at $m\gg \bar\rho V$,
but that it decreases monotonically as $m$ decreases, crossing $0$ 
at $m=\bar\rho V$, and eventually crossing \delv\ at $m$ 
sufficiently smaller than $\bar\rho V$.  In addition, note that 
setting $\delta(t) = -2.71$ and $\delta_c=1.686$ makes 
$\Delta_{\rm NL} \approx 0.2$, implying a void mass of 
$M\approx 0.2\bar\rho V$ for a void of Eulerian volume $V$. 
And finally, thinking of $V$ as a parameter, note that decreasing 
$V$ defines a sequence of nested curves whose limit, as $V\to0$, 
is the constant barrier \delc: $B_{V\to0}(m)\to\delc$.  

The dotted lines in Figure~\ref{eulVoids} show such a nested sequence.
Also shown are two candidate random walks (blue and red solid lines)
which first cross \delv\ at the same Lagrangian mass scale $S=s(M)$,
so that $B_V(M)=\delv$. Since neither of these walks exceeded
\delc\ prior to first crossing \delv, Sheth \& van de Weygaert would
have assigned both walks the same Lagrangian mass and Eulerian
volume. 

For us, the two walks are rather different void candidates.  
This is because the mass inside Eulerian $V$ at time $t$ is given 
by the value of $s(m)$ at which the associated barrier $B_V(m)$ is
first crossed (Sheth 1998).  For the (blue) walk which decreases
monotonically, the monotonicity in Lagrangian $\delta$ translates
directly into a monotonicity in $\Delta_{\rm NL}$, so that conditions
(a), (b) as well as (a$_{\rm L}$), (b$_{\rm L}$) and (c$_{\rm L}$) are
all met. For the void associated with this walk, we would assign the
same mass and volume as would Sheth \& van de Weygaert.  In
particular, the Eulerian volume would lie between $V_3$ and $V_2$.

However, for the other (red) walk, the non-monotonicity of $\delta$
means that $\Delta_{\rm NL}$ is not monotonic either.  More
importantly, although the predicted mass decreases monotonically with
Eulerian $V$, it need not do so smoothly.  Rather, on scales $V$ where
$B_V$ is tangent to the walk, the predicted mass must jump downwards
as $V\to V-\Delta V$ (i.e., as the barrier is made shallower),
because the value of $s$ on which $B_{V-\Delta V}$ is first crossed
can be substantially larger than that on which $B_V$ was first
crossed.  (In the Figure, this happens at about $V_2$.)  
This means that, for the entire portion of the walk
between these two first crossing values (essentially, the value of $s$
at which a barrier $B_V$ is tangent to the walk, and the next larger
value of $s$ at which it pierces the walk), translating the Lagrangian 
$\delta$ to an Eulerian $\Delta_{\rm NL}$ using equation~(\ref{DeltaNL}) 
will {\em not} yield the correct answer.  This, in essence, is why an 
approach based purely on Lagrangian quantities will not work:  
one {\em must} use Eulerian quantities.  
This sharp transition in mass (and hence Eulerian density) at nearly
constant Eulerian volume has a clean physical interpretation in terms
of a dense ``wall'' surrounding the underdense void.
In the current instance, 
we would assign the void an Eulerian volume that is essentially 
$V_2$, with mass interior to the void given by the value of $s$ at 
which $B_{V_2}$ intersects the walk.  And we would interpret the value
of $s$ at which $B_{V_2}$ was tangent to the walk as the mass at the 
void wall.  

For this particular walk, the two masses can be quite different 
indicating that the Eulerian void should be rather well delineated 
by the surrounding Eulerian overdensity.  This is precisely the type 
of void that is easiest to identify observationally -- so it is 
worth noting that it is for just such voids that our algorithm 
can differ substantially from that of Sheth \& van de Weygaert.  
The voids on which we would agree are those associated with walks 
that are similar to the monotonically decreasing walk in 
Figure~\ref{eulVoids}.   Since these correspond to voids for which 
there is no obvious defining `wall', they are hardest to define 
observationally.  

To highlight how different our algorithm is, it is worth contrasting 
the role played by \delv\ in the two approaches.  In the old one, 
first crossings of \delv\ are fundamental, because they give the 
superset of Lagrangian void candidates from which one discards 
those which first crossed \delc, on the basis that they represent voids 
that would have been crushed out of existence by Eulerian evolution.  
One might have thought that, because it accounts for the squeezing 
rather than complete crushing of these regions due to Eulerian 
evolution, our modification mainly serves to reduce the predicted 
volumes of the ones which remain.  While this is correct, there 
is a subtlety.  

As Figure~\ref{eulVoids} shows, if the first crossing of \delv\ 
happens to lie in a region where the $\delta-\Delta_{\rm NL}$ mapping 
of equation~(\ref{DeltaNL}) does not apply, then it is simply not as 
important as subsequent crossings of \delv.  E.g., suppose the spike 
in the walk were higher, so that it crossed above $B_{V_2}$ for a 
while, before dropping down to and zig-zagging around \delv\ a few 
times.  Then the Eulerian region just within $V_2$ would not be a 
void (because the walk crossed $B_{V_2}$ above \delv), but one of 
the subsequent zig-zags around \delv\ might actually be the one 
which first crosses an Eulerian $B_V$, and so represents a squeezed 
Eulerian void.  This one would certainly have a smaller volume than 
that given to the initial first crossing candidate by 
Sheth-van de Weygaert, but clearly, although \delv\ plays an important 
role, the first crossing of \delv\ is not necessarily the most relevant 
one.  
The fact that the first crossing of \delv\ is no longer so important 
is one reason why we have been unable to derive an analytic expression 
for the distribution of void volumes associated with our new formulation 
of the void-in-cloud problem.  We discuss this further in the Appendix.  
It is, of course, straightforward to implement our algorithm numerically, 
and we describe this in the next section.  

But before we do so, we note that our new approach helps alleviate one 
unphysical feature of the old model.  Namely, in the 
Sheth-van de Weygaert approach, the volume fraction covered by 
voids is $5 \delc/(\delc+|\delv|)$.  Since $\delc\approx 1.686$ 
and $|\delv|\approx 2.71$, this `fraction' is nearly 2.  It is easy to 
see that this fraction must be smaller in our new approach, because 
we would assign a smaller Eulerian volume to each of the Sheth-van de 
Weygaert void candidates (in some cases, this volume is vanishingly 
small).  We show below that the associated void covering fraction is 
$1.17$; i.e., although it is still greater than unity, the problem is 
now 20\% rather than 100\%.

\subsection{Correlated vs uncorrelated steps}
We expect our model predictions to depend on whether or not the steps 
in the random walk are correlated.  For walks with uncorrelated steps, 
the solution to the two-barrier Lagrangian void-in-cloud problem 
\delv-\delc\ is quite different from that for the single \delv\ barrier 
void-in-void problem; it has far fewer small voids 
(Sheth \& van de Weygaert 2004).  
We expect our purely Eulerian void-in-cloud algorithm to produce even 
smaller voids, so that all three estimates of the void distribution 
should differ substantially from one another.  

However, we expect these three estimates of void abundances to be 
rather similar for walks with correlated steps.  This is because 
correlated steps generally result in smoother walks.  
Indeed, Paranjape et al. (2011) have recently shown that the limiting 
case of completely correlated steps, 
in which the walk height on one scale $S$ completely specifies its 
height on all other $s$ via $\delta(S)/\sqrt{S} = \delta(s)/\sqrt{s}$, 
actually provides a useful way of thinking about the single barrier 
problem.  
In this limit, walks do not zig-zag at all, which, in the present 
context means that the void-in-cloud problem {\em never} arises, 
so the solution to the single barrier case \delv\ would be the same 
as that for the purely Lagrangian (Sheth-van de Weygaert) formulation 
of the void-in-cloud problem (since no walks will have crossed \delc\ 
prior to crossing \delv).  
Since our algorithm is basically the same as Sheth-van de Weygaert 
for smooth monotonically decreasing walks, the prediction associated 
with our Eulerian void-in-cloud formulation would also reduce to the 
first crossing distribution for the single barrier of height \delv.  
We expect to see differences between these three cases as we move away 
from the completely correlated limit.  But since Paranjape et al. have 
already shown that this limit is essentially exact for the most massive 
objects, we only expect to see differences for low mass voids.  
Walks with uncorrelated steps are far from this limit, so in this case 
we expect to see differences appear at larger masses.  
The next section shows that the predicted importance of the 
void-in-cloud effect does indeed depend on whether or not one accounts 
for correlations between steps.  

\section{Numerical (Monte-Carlo) solution}\label{monteCarlos}
Before we show the numerical Monte-Carlo solution of our algorithm, 
note that although the description above is general, it simplifies 
considerably for power spectra with $P(k)\propto k^n$ with $n=-1.2$.  
In this case, 
 $s\propto m^{-(n+3)/3}\propto m^{-3/5} \sim m^{-1/\delc}$, 
so the barrier shape becomes linear in $s$, and this simplifies the 
numerical analysis considerably.  For this reason, we have chosen to 
present results for this case first.  We show results for a CDM power 
spectrum at the end of this section.  

\begin{figure}
 \centering
 \includegraphics[width=\hsize]{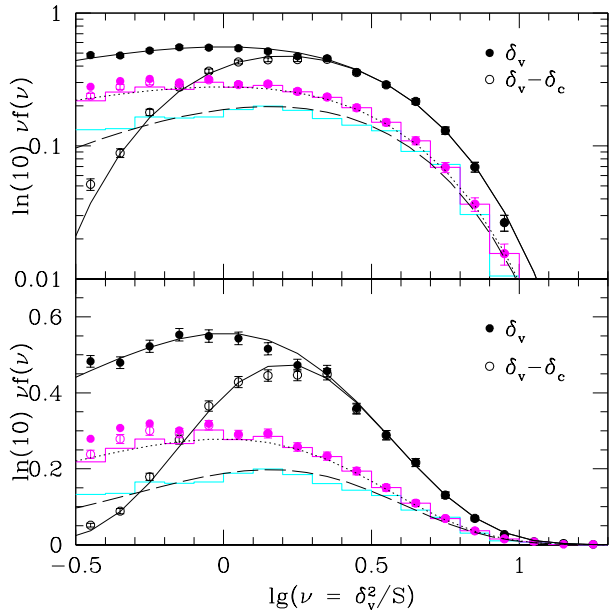}
 \caption{Monte-Carlo solution of various Excursion Set based predictions 
          for void abundances.  
          Filled circles show the first crossing distribution of a 
          single barrier of height \delv; open circles show the 
          distribution of the subset of walks which did not first 
          cross \delc; histograms show the distribution associated 
          with our new algorithm.  
          The black symbols and cyan histogram are for walks with
          uncorrelated steps, while the magenta symbols and histogram 
          (which lie very close to each other)
          are for walks with correlated steps (see text for details).
          (For clarity, we only show error bars for the open circles 
          in this case.)
          The differences between the symbols and the corresponding
          histograms are much more pronounced for walks with
          uncorrelated steps.  Solid curves show the corresponding 
          analytic solutions for walks with uncorrelated steps, for
          the single barrier and two-constant barrier cases. 
          Dashed curve shows 2 times the distribution $Sf_0(S)$
          derived in the Appendix (equation~\ref{fzero-final}), and
          provides an excellent description of the cyan histogram.
          Dotted curve shows the expected solution for walks with 
          completely correlated steps, which describes our results 
          for correlated steps rather well. }
 \label{vfveul}
\end{figure}

Our Monte-Carlo algorithm works as follows.  
For a walk with uncorrelated steps (corresponding to a filter that 
is sharp in $k$-space), we accumulate independent Gaussian random
numbers $g_i$ with a fixed variance
 $\Delta s$, $\del_j^{({\rm uncorr})} = \sum_{i=1}^jg_i$, 
and record the step at which the barrier \delv\ was first crossed as 
well as the step at which \delc\ was first crossed.  
The distribution of $s$ at which \delv\ was first crossed represents 
the solution to the void-in-void problem; that of the subset of walks 
for which \delv\ was first crossed prior to ever crossing \delc\ 
represents the Sheth-van de Weygaert algorithm.  
The black filled and open symbols in Figure~\ref{vfveul} show these 
two distributions, respectively:  the solid black curves going through
them show the associated analytic expressions (from Bond et al. 1991
and  Sheth \& van de Weygaert 2004; equation~\ref{zigzags}).  
The agreement indicates that the numerical algorithm works. 

The cyan histogram shows the result of implementing our algorithm 
as follows.  For a walk that crossed \delv\ at least once, we choose 
all steps prior to the first crossing.  At each step $j$ we have a 
pair $(\delta_j,s_j)$ which together define an Eulerian volume $V_j$.
(In more detail, the value $s_j$ gives a mass $m_j$, and insertion 
of $\delta_j$ in equation~(\ref{DeltaNL}) yields $m_j/\bar\rho V_j$.)  
We call the smallest value of $V_j$ associated with the walk 
so far $V_{\rm min}$.  If $V_{\rm min} = 0$, we stop -- this would 
only have happened if the walk exceeded \delc.  Since this would 
mean the void candidate has been crushed out of existence, we 
eliminate the walk from the list of void-walks.  
If $V_{\rm min}>0$, then asking that it be a void sets a mass 
$M_{\rm min} \approx 0.2\bar\rho V_{\rm min}$, which determines an 
$S_{\rm min}$.  (Typically, this value is larger than that on which 
the walk first crossed below \delv.)  So we check if the walk remains 
below the barrier $B_{V_{\rm min}}(m)$ (of equation~\ref{BVm}) for all 
$s(m)<S_{\rm min}$.  If it does, we store this value and proceed to the 
next walk.  If it does not, then we select the first of all steps 
larger than $S_{\rm min}$ which are below \delv, and repeat the algorithm 
above until a void is identified, or until $S_{\rm min}$ becomes 
sufficiently large that the associated void size is negligibly small.  

In the large mass (or volume) regime where the \delv\ 
(no void-in-cloud) and \delv-\delc\ (Lagrangian void-in-cloud) 
distributions are similar, our algorithm predicts about a factor of 
2 fewer voids.  On smaller scales, where the \delv-\delc\ prediction 
is dropping sharply, ours predicts more voids -- though it is still 
about a factor of 3 smaller than when the void-in-cloud problem has 
been ignored altogether.  This quantifies the discussion at the 
end of the previous section about the expected differences between 
these three ways of estimating void abundances.  

It turns out to be interesting to classify the voids identified by 
our new algorithm in terms of the number of times we had to loop 
through the algorithm.  This is because, in the Appendix, we describe 
an analytic estimate of the fraction of walks $f_0(S)$ for which a
void is identified after only a single pass through the algorithm.  This 
estimate is in good agreement with the fraction of such walks in our 
Monte Carlos (not shown).  Curiously, multiplying this analytic
estimate (equation~\ref{fzero-final}) by a factor of 2 provides an
excellent description of the full set of void walks. This is shown as
the dashed line in \fig{vfveul}.  We have not found a simple derivation 
of why this should have been the case.  Integrating $2f_0(S)$ numerically
over all $S$ gives $0.234$ as the Lagrangian volume fraction; this is in 
excellent agreement with our Monte-Carlos.  (The corresponding Eulerian 
volume fraction is a factor $5$ larger, i.e. $1.17$, as mentioned 
previously.)

The magenta symbols and histogram show the corresponding results for 
walks with correlated steps.  In practice, we transformed each walk 
with uncorrelated steps into one with correlations by applying 
smoothing filters of different scales following Bond \etal\ (1991): 
we apply the filter $W(kR_{\rm L})$ to the same set of numbers $g_i$ 
as above to get
 $\del_j^{\rm(corr)}=\sum_ig_iW(k_iR_{{\rm L}j})$. 
Here $R_{\rm L}$ is the Lagrangian length scale related to mass $m$ by 
$m=(4\pi/3)\bar\rho R_{\rm L}^3$. In this case, the correlation depends 
on the form of the filter and on the shape of the initial linear theory 
power spectrum $P(k)$, since one needs to know which values of $k_j$ 
and $R_{{\rm L}j}$ to associate with the $j$-th step.
Once a power spectrum and filter are specified, this can be done by
inverting the relations
 $j\Delta s = (2\pi^2)^{-1}\int_0^{k_j} {\rm d}k \,k^2 P(k)$ 
and
 $j\Delta s = (2\pi^2)^{-1} \int_0^{\infty}{\rm d}k \,k^2 P(k) W^2(k R_{{\rm L}j})$.
We used a Gaussian smoothing filter $W(kR)={\rm e}^{-(kR)^2/2}$ and
$P(k)\propto k^{-1.2}$.  We then subjected each correlated walk to the
same analysis as for the uncorrelated walks.  


Notice that, in contrast to when the steps were uncorrelated, 
now the three ways of estimating void abundances all give almost 
the same answer.  The sense of the differences which are beginning 
to appear at small masses is easily understood:  ignoring the 
void-in-cloud problem altogether over-estimates the abundances 
relative to the Lagrangian void-in-cloud treatment.  
This is only a small effect because a correlated walk which crosses 
$\delv$ is much less likely to have crossed $\delc$ than an 
uncorrelated walk. Stated differently: most walks which crossed \delv\ 
didn't go into the disallowed ($>\delc$) region anyway, so removing 
them makes little difference.  In turn, the Lagrangian void-in-cloud 
analysis slightly overestimates the abundances relative to our 
Eulerian void-in-cloud algorithm, because it only eliminates the 
voids that got completely crushed, but does not alter the sizes of 
those that got squeezed a little.  
Therefore, it tends to overestimate the sizes of the voids, but 
this only becomes a significant effect for rather small voids. 

\begin{figure}
 \centering
 \includegraphics[width=\hsize]{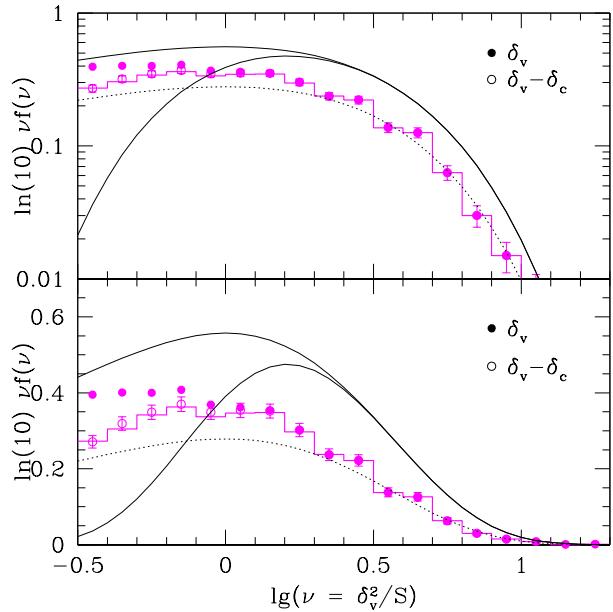}
 \caption{Monte-Carlo distributions for walks with correlated steps
   with the same format as in \fig{vfveul}, except that now
   Monte-Carlos used a $\Lambda$CDM $P(k)$ and Tophat smoothing
   filters. The solid and dotted curves are the same as in \fig{vfveul}.}  
 \label{vfvcdm}
\end{figure}

By a curious coincidence, the first crossing distribution for walks 
with ``completely correlated'' steps in the presence of a single 
constant barrier of height \delv\ 
\be
 sf_v^{\rm (cc)}(s) = \frac12\frac{|\delv|}{\sqrt{2\pi s}}\,
                           {\rm e}^{-\delv^2/2s} \,, 
 \label{compcorr}
\ee
(Paranjape \etal\ 2011), 
shown by the dotted curve, provides a rather good description of 
our predicted distribution.  Recall that this prediction assumes 
the walks are perfectly smooth, so there is no void-in-cloud 
problem to begin with.  

First crossing distributions for walks with correlated steps are 
relatively insensitive to shape of the underlying power spectrum 
or the smoothing filter (Bond \etal\ 1991; Paranjape \etal\ 2011).  
This is also true for the problem studied here:  Figure~\ref{vfvcdm} 
shows the result of using $P(k)$ for a flat $\Lambda$CDM model with 
$(\sigma_8,\Omega_m)= (0.8,0.27)$, and Tophat smoothing filters.  
Note that the void-in-cloud problem only becomes noticable for small 
voids -- but that it is more noticable than it was for the Gaussian 
filtered walks (compare filled and open circles here and in previous 
Figure).  In addition, the solution is now noticably different from that 
for `completely correlated' steps.  The sense of both these trends 
is easily understood from the fact that Gaussian smoothing is known to 
produce smoother walks than Tophat (Paranjape \etal\ 2011), so the 
results here are intermediate between those for sharp-$k$ and Gaussian 
smoothing.

\section{Discussion}
We have presented what we believe to be a better Excursion Set 
treatment of the void-in-cloud problem (Section~\ref{constant+linear} 
and Figure~\ref{eulVoids}).  
In addition to accounting for the fact that some voids can be crushed 
completely if they are surrounded by an overdensity which collapses 
around them, our Eulerian-space based approach also accounts for those 
voids which are squeezed rather than completely crushed.  We argued 
that voids which are being squeezed by their surroundings may be the 
easiest to recognize observationally, so our modification is potentially 
an important one.  
In particular, our approach shows explicitly why, in some cases, a 
purely local Lagrangian-based prediction for the evolution yields 
the wrong answer; it can be thought of as an explicit demonstration 
of how stochasticity in the mapping between the Lagrangian and 
Eulerian density fields can arise naturally.  

The Excursion Set statement of the problem involves random walks.  
We provided an analytic expression (Appendix) for the predicted 
distribution of void sizes for the case in which the walks have 
uncorrelated steps, and showed that it was in good agreement with 
numerical Monte-Carlo solutions of the problem.  (This expression 
suffers from a curious `factor-of-two' problem which we discuss 
briefly -- but we leave an exact solution of it to future work.)  
This analysis suggests that the void-in-cloud process modifies the 
predicted void size distribution significantly, so that our new 
treatment of it was necessary.  

However, this conclusion 
depends strongly on whether or not 
we account for the correlations between scales in the initial density 
fluctuation field.  
In contrast to what happens for uncorrelated steps, for correlated 
steps, we found that the change in void abundances due to this effect 
is negligible for voids that are larger than $V_*$, where $V_*$ is the 
characteristic Eulerian scale associated with voids:
  $V_* = 5 M_*/\bar\rho$, where $\sigma(M_*)=|\delta_v|$.  
For flat $\Lambda$CDM with $(\Omega_m,\sigma_8)=(0.27,0.8)$, 
this scale is $V_*\simeq (1.4h^{-1}{\rm Mpc})^3$ at $z\sim 0$; 
larger voids have not been squeezed by their surroundings.  
Since voids identified in most galaxy catalogs are typically much 
larger, it may be unnecessary to account for the void-in-cloud 
process when interpreting observations.  
In this case, the void size distribution is quite well approximated by
that of a single barrier, for which good analytic approximations are  
available (Paranjape et al. 2011).  

Although our results provide increased understanding of void abundances 
and evolution, a number of issues must be addressed before they can be 
used to provide useful constraints on cosmology.  
First, the correlated walk problem is known to underpredict the 
abundances of clusters (Bond \etal\ 1991).  Paranjape \etal\ (2011) 
describe at least three possible resolutions, which have to do with 
the fundamental assumptions which the excursion set approach uses to 
relate the first crossing distribution with halo abundances.  Presumably, 
this problem, and hence the potential resolutions, also apply to voids.  
Second, we must include a model for transforming our knowledge of voids 
in the dark matter distribution to underdensities in the galaxy 
distribution.  This will require applying the analysis of 
Furlanetto \& Piran (2006) to our new formula for void abundances, 
perhaps accounting for the fact that the voids have non-trivial 
internal density profiles (Patiri \etal\ 2006, following Sheth 1998).

\section*{Acknowledgments}
We thank the anonymous referee for helpful comments which signficantly
improved the manuscript.
TYL was supported by the World Premier International Research Center 
Initiative (WPI Initiative), MEXT, Japan and a JSPS travel grant 
(Institutional Program for Young Researcher Overseas Visits).  He is 
grateful to the Center for Particle Cosmology at the University of 
Pennsylvania and to ICTP for their hospitality in 2010 and 2011, 
respectively.  RKS is supported in part by NSF-AST 0908241.

\appendix

\section{The first crossing distribution for walks with uncorrelated steps}

In this appendix we sketch the derivation of the first crossing
distribution $f_0(S)$ (for walks with uncorrelated steps) discussed
in section 3, which counts the fraction of walks that survive
one pass through our algorithm. The analysis is tractable when the
void-in-cloud barriers are linear, which happens for a power spectrum
$P(k)\propto k^{-1.2}$. It is then convenient to think of the barrier
$B_V(m)$ as a function of $s=s(m)$, parametrized by the value
$S=s(M=0.2\bar\rho V)$ at which the barrier crosses the constant
barrier \delv. We use the notation $B_S(s)$ to denote the void-in-cloud 
barrier, and \eqn{BVm} translates to $B_S(s)=\delc-(\delT/S)s$ where
$\delT\equiv \delc+|\delv|$. 

We are after the fraction of walks which satisfy the
following conditions: 
\begin{itemize}
\item They first cross the barrier $B_S(s)$ at $s=S^\prime<S$, without
  having crossed \delv\ before $S^\prime$.
\item They first cross the barrier $B_{S+\Delta S}(s)$ \emph{after}
  this barrier has passed through \delv, i.e., at $s>S+\Delta S$.
\end{itemize}
We evaluate the resulting fraction in the limit $\Delta S\to0$, and
interpret it as $\Delta S f_0(S)$.

The first condition above requires us to compute the fraction of walks
$dS^\prime\Cal{F}_{B}(S^\prime)$ which first cross $B_S(s)$ in the interval
$s\in (S^\prime,S^\prime+dS^\prime)$, without having crossed
\delv\ before. The second condition requires the fraction $ds
f_{B\Delta}(s|S^\prime,B_S(S^\prime))$ of walks that started at height
$B_S(S^\prime)$ on scale $S^\prime$ and then went on to first cross
$B_{S+\Delta S}(s)$ in the range $(s,s+ds)$. The distribution $f_0(S)$
is then given by
\be
\Delta S f_0(S) = \int_0^SdS^\prime \Cal{F}_B(S^\prime) \int_{S+\Delta
S}^\infty ds\,f_{B\Delta}(s|S^\prime,B_S(S^\prime))\,.
\label{fzero-int}
\ee
The distribution $\Cal{F}_B(s)$ can be written in a form which
allows a recursive calculation: we first count the fraction of walks
$f_B(s)$ which first cross $B_S$ at $s<S$, regardless of whether they
crossed \delv, and then subtract those which did cross \delv\ prior to
$s$. We then have
\begin{equation}
 {\cal F}_B(s) = f_B(s) - \int_0^s {\rm d}s'\,{\cal
   F}_v(s')\,f_B(s|s',\delv)\,,
\label{SvdW-2}
\end{equation}
where ${\cal F}_v(s')$ (with $s'<S$) denotes the
distribution of first crossing of \delv\ without crossing $B_S$, and
the integral in the second term is counting walks that reached
\delv\ at $s'$ for the first time without crossing $B_S$, and then
reached $B_S$ for the first time at $s$. A similar argument, with the
roles of $B_S$ and \delv\ interchanged, allows us to write
\begin{equation}
 {\cal F}_v(s') = f_v(s') - \int_0^{s'} {\rm d}s''\,{\cal
   F}_B(s'')\,f_v(s'|B_S,s'') \,,
 \label{SvdW}
\end{equation}
where $f_v(s')$ is the distribution of first crossing of $\delta_v$
whether or not it had first crossed $B_S$, and $f_v(s'|B_S,s'')$ is the
corresponding conditional distribution.
Notice that, despite the compact notation, both the distributions
$\Cal{F}_B(s)$ and $\Cal{F}_v(s)$ depend on the scale $S$ which
parametrizes the barrier $B_S(s)$.
Repeated substitution of ${\cal F}_B(s)$ in the expression for 
${\cal F}_v(s)$, and vice-versa, gives rise to an alternating series, 
the successive terms of which describe walks which cross $\delta_v$ 
after more and more zigs and zags.  

As an aside, we note that when $B_S(s) = \delta_c$ for all $s$,
independent of $S$ (i.e., the Sheth-van de Weygaert model in which
$\delta_c$ is parallel to $\delta_v$), then the integrals in  the
expressions above can be done analytically, yielding
\begin{equation}
 {\cal F}_v(s) = \Fsw(s,\delv,\delc) \equiv f_v(s) - f_{c+T}(s) +
 f_{v+2T}(s) - \ldots  
 \label{zigzags}
\end{equation}
where $c, v$ and $T$ in the expression above denote $\delta_c,
|\delta_v|$  and $\delta_T=\delta_c+|\delta_v|$ (Lam \etal\ 2009;
D'Amico \etal\ 2011). The distribution $\Cal{F}_B$ in this case is
given simply by interchanging the roles of \delc\ and \delv\ in
$\Cal{F}_v$. 
In this case, the first few terms dominate -- walks with many zig-zags 
from one barrier to the other are rare -- so truncating the series 
yields a good approximation to the full answer.  In particular, when 
$\delta_T\gg |\delta_v|$ then the first term dominates for small $S$:
for the largest voids, the void-in-cloud problem is irrelevant.  

However, when $B_S(s)$ is a decreasing function of $s$, which
eventually crosses $\delta_v$, then truncating the series is
dangerous, because close to the point where the two barriers cross,
zig-zags are no longer large, so many can occur. The first crossing
distribution then becomes very sensitive to the exact relation between
the first crossing scale, and the scale at which the barriers
cross. We are interested in the case when $B_S(s)=\delc - (\delT/S)s$
is a linear barrier. The expression for $\Cal{F}_v$ above can then be
solved exactly. Since the full proof involves some rather tedious
integrals, we only present a sketch here highlighting the main
ingredients.  Equations~\eqref{SvdW-2} and~\eqref{SvdW} can be
combined as 
\begin{equation}
\F_v(s') = f_v(s') + \I{1} + \I{2}\,,
\label{uncorr-master}
\end{equation}
where
\begin{align}
\I{1}(s') &= - \int_0^{s'} {\rm d}s\,f_v(s'|s,B_{S}(s))\,f_B(s) \,,
\nonumber\\  
\I{2}(s') &= \int_0^{s'} {\rm d}s_1\,f_v(s'|s_1,B_{S}(s_1))
\nonumber\\ 
&\ph{=+\int_0^S}\times
\int_0^{s_1}{\rm d}s_2f_B(s_1|s_2,\delta_v) \F_v(s_2)\,.
\label{I12-def}
\end{align}
The conditional distributions simplify as in the two-constant barrier
case, due to the Markovian nature of the walks. Unlike this previous
case, however, this time the single barrier distributions $f_v$ and
$f_B$ are fundamentally different from each other. Whereas $f_v$
is the same as before for a single constant barrier, $f_v(s) = (2\pi
s^3)^{-1/2}|\delta_v| e^{-\delta_v^2/2s}$, $f_B$ for a single linear
barrier $B_{S}(s)$ is now given by the Inverse Gaussian (Sheth
1998), 
\be
f_B(s)=f_{\rm IG}(s,B_S(s))\equiv \frac{B_{S}(0)}{(2\pi s^3)^{1/2}}
e^{-B_{S}(s)^2/2s}\,. 
\label{fIG}
\ee
Since the distribution $\F_v(s)$ appears in
the integral in \I{2}, the basic strategy is to recursively use
\eqn{uncorr-master} and solve for $\F_v(s)$. For example, the integral 
in \I{1}\ above reduces to
\begin{align}
\I{1}(s) &= -\frac{(2\delc+|\delv|)}{\sqrt{2\pi s^3}}
e^{-\delv^2/2S} e^{-\frac1{2s}(2\delc+|\delv|)^2(1-s/S)}\,.
\label{I1-eval}
\end{align}
In practice we compute the first few terms of the
recursive series, which clearly reveal a pattern that closely mimicks
the one in equation \eqref{zigzags} above, but with a rescaled
argument for the single barrier distributions. We are left with
\begin{equation}
\F_v(s) = \left(1-\frac{s}{S}\right)^{-3/2}
e^{-\delv^2/2S}
\Fsw\left(\frac{s}{1-s/S},\delv,\delc\right)\,, 
\label{SvdW-rescaled}
\end{equation}
where $\Fsw(t,\delv,\delc)$ was defined in \eqref{zigzags}. 
For the interested reader, the integrals appearing in the derivation
of equation \eqref{SvdW-rescaled}  involve repeated use of the
identities (for $A,B>0$)  
\begin{align}
\int_0^1 \frac{dy}{y^{3/2}}\frac1{\left(1-y\right)^{3/2}}
e^{-\frac{A^2}{2y}- \frac{B^2}{2(1-y)}} &= \sqrt{2\pi} \frac{A+B}{AB}
e^{-\frac12(A+B)^2} \,,\nonumber\\
\int_0^1 \frac{dy}{y^{1/2}}\frac1{\left(1-y\right)^{3/2}}
e^{-\frac{A^2}{2y}- \frac{B^2}{2(1-y)}} &= \frac{\sqrt{2\pi}}{B}
e^{-\frac12(A+B)^2} \,,
\label{integ-ids}
\end{align}
which can be proved using the relation 3.472(5) of Gradshteyn \&
Rhyzik (2007).

Using \eqn{SvdW-rescaled} in \eqn{SvdW-2} leads to an expression for
$\Cal{F}_B(S^\prime)$, which can then be used in \eqn{fzero-int} to
obtain an expression for $f_0(S)$, after substituting for the
conditional distribution $f_{B\Delta}(s|S^\prime,B_S(S^\prime))$ using
\be
f_{B\Delta}(s|S^\prime,B_S(S^\prime)) = f_{\rm
  IG}(s-S^\prime,B_{S+\Delta S}(s)-B_S(S^\prime))\,,
\label{fBDelta}
\ee
which follows from the Markovianity of the walks, with $f_{\rm IG}$
defined in \eqn{fIG}. Notice that, in the limit $\Delta S\to 0$, the
evaluation of $B_{S+\Delta S}(s)-B_S(S^\prime)$ at $s=S^\prime$
becomes proportional to $\Delta S$, and hence the remaining expression
for $f_0(S)$ can be evaluated at $\Delta S=0$. Unfortunately, not all
of the resulting integrals can be performed analytically. It turns out
to be better to use the series for \Fsw\ given in equation~(1) of 
Sheth \& van de Weygaert (2004),  
\begin{equation}
 \Fsw(t,\delv,\delc) = \sum_{j=1}^\infty \frac{j\pi}{\delT^2}
                       \,\sin\left(\frac{j\pi|\delv|}{\delT}\right)
                       \,{\rm e}^{-\frac{j^2\pi^2}{2\delT^2}t}\,,
\label{SvdW-original}
\end{equation}
which is equivalent to the one in equation \eqref{zigzags} (D'Amico
\etal\ 2011). This allows us to bring the expression for $f_0$ to the
following form 
\be
Sf_0(S) = \frac1{\sqrt{2\pi}}e^{-\nuv^2/2}\bigg[ \Cal{A} - \Cal{B}
  \bigg]\,, 
\label{fzero-final}
\ee
where
\begin{align}
\Cal{A} &= \nuc\nuT e^{\nuc^2/2}\left(\sqrt{2\pi}\,
\erfc{\frac{\nuc}{\sqrt{2}}} - \nuT e^{\nuT^2/2}\,{\rm
  I}(\nuc,\nuT)\right)\,,  
\label{fzero-A}\\
\Cal{B} &= \sum_{j=1}^\infty j\pi
\sin\left(\frac{j\pi|\delv|}{\delT}\right)
\bigg[ \frac{4\nuT}{j^2\pi^2} - \frac2{\nuT}e^{j^2\pi^2/2\nuT^2} 
  \Gamma\left(0,\frac{j^2\pi^2}{2\nuT^2}\right) \nonumber\\
&\ph{\sum_{j=1}^\infty j\pi}
 +\sqrt{2\pi}\,{\rm II}(j,\nuT) -\nuT \,{\rm III}(j,\nuT) \bigg]\,,
\label{fzero-B}
\end{align}
and we used the notation $\{\nuv,\nuc,\nuT\} \equiv
\{|\delv|,\delc,\delT\}/\sqrt{S}$ and defined the integrals
\begin{align}
{\rm I}(\nuc,\nuT) &= \int_0^1\frac{dy}{\sqrt{y}}
\exp\left(-\frac{\nuT^2y}{2}-\frac{\nuc^2}{2y}\right) \nonumber\\ 
&\ph{\int_0^1\frac{dy}{\sqrt{y}}\exp\left(\right)}
\times \erfc{\frac{\nuT}{\sqrt{2}}\sqrt{1-y}} \,,  
\label{fzero-I}
\end{align}
\begin{align}
{\rm II}(j,\nuT) &= \int_0^1\frac{dy}{\sqrt{y}}
\exp\left(-\frac{\nuT^2y}{2} -\frac{j^2\pi^2(1-y)}{2\nuT^2y} \right)
\nonumber\\ 
&\ph{\int_0^1\frac{dy}{\sqrt{y}}\exp\left(\right)}
\times\erfc{\frac{\nuT\sqrt{y}}{\sqrt{2}}} \,,
\label{fzero-II}
\end{align}
\begin{align}
{\rm III}(j,\nuT) &= \int_0^1\frac{dx}{x}
\exp\left(-\frac{j^2\pi^2(1-x)}{2\nuT^2x} \right) \nonumber\\
&\ph{\int_0^1}
\times \int_0^1 \frac{dy}{(1-y)^{3/2}}(1-xy) \exp
\left(-\frac{\nuT^2x}{2} \frac{y^2}{1-y} \right) \nonumber\\
&\ph{\int_0^1\frac{dy}{\sqrt{y}}\exp\left(\right)}
\times\erfc{\frac{\nuT\sqrt{xy}}{\sqrt{2}}} \,,
\label{fzero-III}
\end{align}
which must be performed numerically. We have checked that the result
(after keeping $\sim600$ terms in the sum over $j$) accurately
describes the fraction of walks which survive one pass through our
numerical algorithm, as it should. 

As noted in the text, we find that 2 times $f_0(S)$ provides 
an excellent description for the full Monte Carlo distribution 
obtained by our algorithm. This is shown as the dashed line in~\fig{vfveul}. 
In fact, if we write this factor of two as follows, 
\begin{equation}
 Sf(S) = 2\,Sf_0(S) = \frac{Sf_0(S)}{1-1/2} = Sf_0(S)
 \sum_{n=0}^\infty 2^{-n}\,, 
\label{fzero}
\end{equation}
then the $n$th term in the sum above approximates the distribution 
associated with the set of walks which required $n+1$ loops through 
our algorithm.  While this is highly suggestive of a resummation of 
``loops'', with the ``tree-level'' result being $f_0(S)$, we have 
not found a simple demonstration of why this should be the case.

\label{lastpage}

\end{document}